\documentclass[pra,aps,twocolumn,showpacs,showkeys]{revtex4}
\usepackage{times}
\usepackage{graphicx}
\usepackage{subfigure}
\usepackage{amsmath}
\providecommand{\ignore}[1]{}
\begin{document}

\title{Error Analysis For Encoding A Qubit In An Oscillator}
\author{S. Glancy}
\email{sglancy@boulder.nist.gov}
\author{E. Knill}
\email{knill@boulder.nist.gov}
\affiliation{Mathematical and Computing Science Division, Information Technology Laboratory, National Institute of Standards and Technology, Boulder, Colorado 80301, USA}
\date{\today}
\begin{abstract}
In the paper titled ``Encoding A Qubit In An Oscillator'' Gottesman, Kitaev, and Preskill [quant-ph/0008040] described a method to encode a qubit in the continuous Hilbert space of an oscillator's position and momentum variables. This encoding provides a natural error correction scheme that can correct errors due to small shifts of the position or momentum wave functions (i.e., use of the displacement operator). We present bounds on the size of correctable shift errors when both qubit and ancilla states may contain errors. We then use these bounds to constrain the quality of input qubit and ancilla states.
\end{abstract}
\maketitle

Most quantum computation schemes propose encoding qubits using natural
two level systems such as spin $1/2$ particles. Others exploit only
two states of a larger discrete Hilbert space such as the energy
excitation levels of an ion.  In the paper ``Encoding A Qubit In An
Oscillator'' \cite{Gottesman01}, Gottesman, Kitaev, and Preskill (GKP)
described an alternative method for encoding a qubit in the
continuous position and momentum degrees of freedom of an oscillator.
Because the qubit is encoded in an infinite dimensional Hilbert space,
a natural error correcting scheme arises.  It can correct errors due
to shifts in the oscillator's position or momentum ({\em i.e.}, small
application of the displacement operator).  When the oscillator is
chosen to be a single mode of the electromagnetic field, fault
tolerant computation can be performed by use of only phase shifters, beam
splitters, squeezing, photon counting, and homodyne
measurements. \cite{Gottesman01, Bartlett02}.  However, state
preparation requires nonlinear interactions.

The GKP scheme constitutes a type of linear optical quantum computer.
Other schemes for such computers are based on
the proposals of Knill, Laflamme, and Milburn (KLM) \cite{Knill00a,
Knill00b, Knill01} and of Ralph {\em et al.} \cite{Ralph03}.
Of these, the KLM proposal has received most of the analysis
\cite{Knill00b, Glancy02, Silva04, Silva05, Dawson05}. However,
which (if any) of these three schemes is most suitable for large scale
quantum computation is still unknown. Thus, further analysis of all
three schemes is required.  This paper is devoted to an analysis of
the GKP quantum computer.  We consider errors only in the input
qubit and ancilla states, and we obtain a threshold for the maximum
size of position and momentum shifts that the error correcting scheme
can always correct. We then use this threshold to constrain the
quality of the input states and estimate the number of photons the
input states must contain.

We first give a brief review of qubits in the GKP scheme.  One may use
any type of oscillator to represent a qubit, but for the purposes of
this paper we choose the oscillators to be single modes of the
electromagnetic field.  Let $\hat{a}_i$ be the photon annihilation
operator of mode $i$, $\hat{a}_i^\dagger$ the creation operator,
$\hat{x}_i=1/\sqrt{2}(\hat{a}_i^\dagger+\hat{a}_i)$ the
$x$-quadrature, and $\hat{p}_i=i/\sqrt{2}(\hat{a}_i^\dagger-\hat{a}_i)$
the $p$-quadrature. Let $|x\rangle$ denote eigenstates of $\hat{x}$, and
$|p\rangle$ eigenstates of $\hat{p}$.  We represent the logical 0
qubit as
\begin{eqnarray}
|0_\text{L}\rangle & = & \sum_{s=-\infty}^{\infty}\delta(x-2s\sqrt{\pi})|x\rangle \\
 & =& \frac{1}{\sqrt{2}}\sum_{s=-\infty}^{\infty}\delta(p-s\sqrt{\pi})|p\rangle,
\end{eqnarray}
and the logical 1 qubit is
\begin{eqnarray}
|1_\text{L}\rangle & = & \sum_{s=-\infty}^{\infty}\delta(x-(2s+1)\sqrt{\pi})|x\rangle \\
 & = & \frac{1}{\sqrt{2}}\sum_{s=-\infty}^{\infty}(-1)^s\delta(p-s\sqrt{\pi})|p\rangle.
\end{eqnarray}
%
%\begin{eqnarray}
%|0_\text{L}\rangle & = & \sum_{s=-\infty}^{\infty}\delta(x-2s\alpha)|x\rangle %\\
% & =& %\frac{\sqrt{2\pi}}{2\alpha}\sum_{s=-\infty}^{\infty}\delta(p-\frac{s\pi}{\alpha%})|p\rangle,
%\end{eqnarray}
%and the logical 1 qubit will be
%\begin{eqnarray}
%|1_\text{L}\rangle & = & %\sum_{s=-\infty}^{\infty}\delta(x-(2s+1)\alpha)|x\rangle \\
% & = & %\frac{\sqrt{2\pi}}{2\alpha}\sum_{s=-\infty}^{\infty}(-1)^s\delta(p-\frac{s\pi}{%\alpha})|p\rangle.
%\end{eqnarray}
%
In the $x$-quadrature these states are infinitely long combs of delta functions, with the $|1_\text{L}\rangle$ displaced a distance $\sqrt{\pi}$ from the $|0_\text{L}\rangle$ state.  Such states are clearly unphysical, having an infinite energy expectation value, and they are not normalizable.

This encoding can protect the qubits from errors due to displacements in the $x$- and $p$-quadratures, described by the operators $e^{-iu\hat{p}}$ (shifting the $x$-quadrature a distance of $u$) and $e^{-iv\hat{x}}$ (shifting the $p$-quadrature a distance of $v$).  These shift operators form a complete basis, so any error superoperator $\mathcal{E}$ acting on the density operator $\rho$ of a single oscillator may be expanded as \cite{Gottesman01}
\begin{eqnarray}
\mathcal{E}(\rho)=\int\mathrm{d}u\mathrm{d}v\mathrm{d}u'\mathrm{d}v' F(u,v,u',v') e^{-iu\hat{p}}e^{-iv\hat{x}}\rho e^{iu'\hat{p}}e^{iv'\hat{x}}.
\end{eqnarray}
If the distribution $F(u,v,u',v')$ is sufficiently concentrated near the origin, then this encoding allows us to correct $\mathcal{E}$ and recover $\rho$.

Measurement of the qubits can be accomplished by use of homodyne detection of the $x$-quadrature.  Results in which the $x$-quadrature is measured to be closer to an even multiple of $\sqrt{\pi}$ are registered as the detection of $|0_\text{L}\rangle$, and measurements closer to an odd multiple are registered as $|1_\text{L}\rangle$.  Therefore, a shift error that is larger than $\sqrt{\pi}/2$ may cause an error in the qubit measurement.

\begin{figure}
\begin{flushleft}
(a)
\end{flushleft}
\begin{center}
\includegraphics[width=80mm]{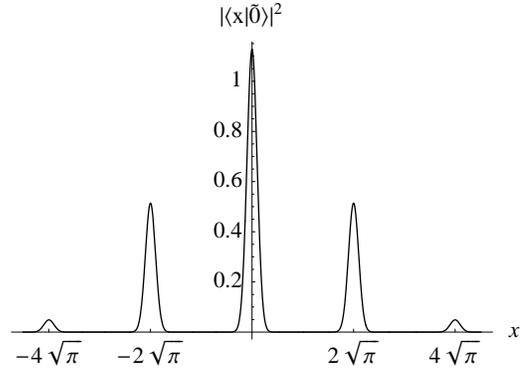}\\
\begin{flushleft}
(b)
\end{flushleft}
\includegraphics[width=80mm]{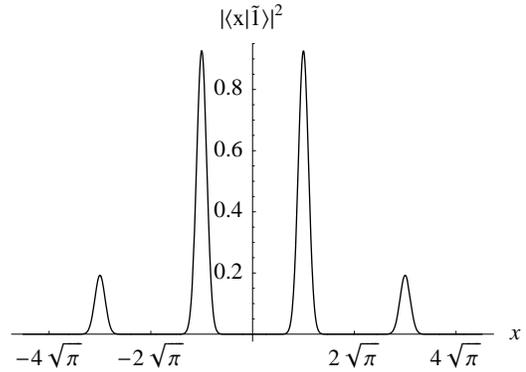}
\end{center}
\caption{Example approximate qubit states. (a) shows the $x$-quadrature wave function of $|\tilde{0}\rangle$, and (b) shows $|\tilde{1}\rangle$.  For each of these we have chosen $\Delta=k=1/4$ \label{approx_qubits}}
\end{figure}

Because the qubit states are unphysical, we must find some states that can approximate the ideal qubit states and are physically realizable.  GKP propose using states whose $x$-quadrature wave function is composed of a series of Gaussian peaks of width $\Delta$ contained in a larger Gaussian envelope of width $1/k$.  The approximations of the 0 and 1 logical states are
\begin{equation}
\langle x |\tilde{0}\rangle = N_0\sum_{s=-\infty}^{\infty}e^{-\frac{1}{2}(2sk\sqrt{\pi})^2}e^{-\frac{1}{2} \left(\frac{x-2s\sqrt{\pi}}{\Delta} \right)^2}
\end{equation}
and
\begin{equation}
\langle x |\tilde{1}\rangle = N_1\sum_{s=-\infty}^{\infty}e^{-\frac{1}{2}((2s+1)k\sqrt{\pi})^2} e^{-\frac{1}{2} \left(\frac{x-(2s+1)\sqrt{\pi}}{\Delta} \right)^2},
\end{equation}
where $N_0$ and $N_1$ are normalization factors. In the limit that $k$
and $\Delta$ are both small, $N_0\approx N_1=\sqrt{2k}$. We show
examples of these approximate qubit states in
Fig. \ref{approx_qubits}. First, note that these states are not
orthogonal, and there will be some probability of mistaking a 0 state
for a 1 state (and a 1 for a 0).  This probability is equal to the
probability that the $x$-quadrature of the 0 state is measured to be
closer to an odd multiple of $\sqrt{\pi}$.
\begin{equation}
P_{0\rightarrow 1} = \sum_{n=-\infty}^{\infty} \int_{2\sqrt{\pi}n+\sqrt{\pi}/2}^{2\sqrt{\pi}(n+1)-\sqrt{\pi}/2} \text{d}x |\langle x|\tilde{0}\rangle|^2.
\end{equation}
In the limit of small $\Delta$ and $k$, this expression can be approximated as
\begin{eqnarray}
P_{0\rightarrow 1} & \approx & \frac{2}{\sqrt{\pi}} \sum_{n=0}^{\infty} \int_{\frac{2\sqrt{\pi}n+\sqrt{\pi}/2}{\Delta}}^{\frac{2\sqrt{\pi}(n+1)+\sqrt{\pi}/2}{\Delta}} e^{-y^2} \text{d}y \\
 & \approx & \frac{2}{\sqrt{\pi}} \int_{\frac{\sqrt{\pi}}{2\Delta}}^{\infty} e^{-y^2} \text{d}y.
\end{eqnarray}
For $\Delta=k=0.5$, $P_{0\rightarrow 1} \sim 0.01$, and for
$\Delta=k=0.25$, $P_{0\rightarrow 1} \sim 10^{-6}$.  For small $\Delta$ and $k$, $P_{1\rightarrow 0}\approx P_{0\rightarrow 1}$. However, we desire not
only that measurements can distinguish between the 0 and 1
states, but also that an error correcting procedure can reliably
correct errors.  This will place stricter requirements on the
approximate qubit states.

%We can express these states in the $p$-quadrature through the Fourier transform:
%\begin{equation}
%\langle p|\tilde{0}\rangle= \frac{N_0\sqrt{\Delta}\pi^{1/4}}{\sqrt{2}k\alpha}\sum_{s=-\infty}^{\infty}e^{-\frac{1}{2}\left(\frac{s\pi}{\alpha}\right)^2\frac{\Delta^2}{\Delta^2k^2+1}}e^{-\frac{1}{2}\left(\frac{p-\frac{s\pi}{\alpha\left(\Delta^2k^2+1\right)}}{\frac{k}{\sqrt{\Delta^2k^2+1}}}\right)^2}
%\end{equation}

\begin{figure}
\includegraphics[width=8cm]{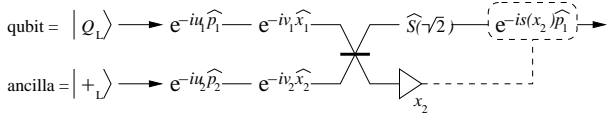}
\caption{Procedure to correct $x$ shifts acting on the qubit.  The circuit shows errors acting on the qubit and ancilla.  Then the qubit and ancilla modes meet in a beam splitter, mode 1 is squeezed, the $x$-quadrature of mode 2 is measured, and a correcting shift of $s(x_2)$ is applied to the qubit. \label{correcting_x}}
\end{figure}

We now examine the error correcting circuits.  We imagine that the
qubit $|Q_L\rangle$ initially exists in a superposition of the
$|0_\text{L}\rangle$ and $|1_\text{L}\rangle$ states in mode 1.  The
qubit then receives errors resulting in shifts of $u_1$ in $x_1$ and
$v_1$ in $p_1$.  We begin the correction procedure by repairing the
shifts in $x_1$.  This requires an ancilla qubit prepared in the state
$|+_\text{L}\rangle=|0_\text{L}\rangle+|1_\text{L}\rangle$ in mode
2. Of course, the ancilla may also be subject to errors, resulting in
shifts $u_2$ and $v_2$.  After the errors, the state of the system is
\begin{equation}
e^{-iu_1\hat{p}_1}e^{-iv_1\hat{x}_1}e^{-iu_2\hat{p}_2}e^{-iv_2\hat{x}_2}|Q_{\text{L}}\rangle_1|+_{\text{L}}\rangle_2.
\end{equation}
To correct the qubit's errors, modes 1 and 2 are sent through the network pictured in Fig. \ref{correcting_x}. The two damaged modes meet in a beam splitter, which performs the transformation
\begin{eqnarray}
x_1 & \rightarrow & \frac{1}{\sqrt{2}}\left(x_1 - x_2\right) \\
x_2 & \rightarrow & \frac{1}{\sqrt{2}}\left(x_1 + x_2\right)
\end{eqnarray}
in the $x$-quadrature wave function of the two modes.  After mode 1 exits the beam splitter, we apply the squeezing operator $\hat{S}(\sqrt{2})$, where $\hat{S}$ is defined by $\hat{S}(q)|x\rangle=\sqrt{q}|x/q\rangle$.  We now measure the $x$-quadrature of mode 2 and obtain the result $x_2=1/\sqrt{2}\left(n\sqrt{\pi}-u_1-u_2 \right),$ where $n$ may be any integer.  This measurement provides some information about the errors shifting the $x$-quadrature of the qubit.  The errors can be (partially) corrected by applying the displacement operator $e^{-is(x_2)\hat{p}_1}$, where
\begin{equation}
s(q)=-\frac{q}{\sqrt{2}}+\frac{1}{2}\text{Mod}_{2\sqrt{\pi}}\left(2\sqrt{2}q \right).
\end{equation} 
Here the Mod function has the range $[-\sqrt{\pi},\sqrt{\pi})$.

This entire sequence of operations -- errors and correction procedure --  transforms the qubit to the state
\begin{eqnarray}
|Q_\text{L}\rangle & \rightarrow & e^{i\phi(u_1,v_1,u_2,v_2,n)} e^{-i(v_1-v_2)\hat{x}_1} \nonumber \\
& & \times e^{-i\left(u_1-\frac{1}{2}\text{Mod}_{2\sqrt{\pi}}\left(2u_1+2u_2 \right) \right)\hat{p}_1
}|Q_\text{L}\rangle,
\end{eqnarray}
where $e^{i\phi(u_1,v_1,u_2,v_2,n)}$ is a phase factor independent of the input qubit.  If $|u_1+u_2|<\sqrt{\pi}/2$, then the qubit is ``corrected'' to the state
\begin{eqnarray}
|Q_\text{L}\rangle & \rightarrow & e^{i\phi(u_1,v_1,u_2,v_2,n)} e^{-i(v_1-v_2)\hat{x}_1} \nonumber \\
& & \times e^{-iu_2\hat{p}_1}|Q_\text{L}\rangle.
\label{x_correction}
\end{eqnarray}
Notice that both the ancilla and the qubit's $p$ shift errors $v_1$
and $v_2$ both appear now acting on the qubit. Also, the qubit has
been affected by the ancilla's $x$ shift error $u_2$. In the case when
$|u_1+u_2|>\sqrt{\pi}/2$, the error correcting procedure applies the
Pauli $X$ operator to the qubit state, producing
\begin{eqnarray}
|Q_\text{L}\rangle & \rightarrow & e^{i\phi(u_1,v_1,u_2,v_2,n)} e^{-i(v_1-v_2)\hat{x}_1} \nonumber \\
& & \times e^{-iu_2\hat{p}_1}X|Q_\text{L}\rangle.
\end{eqnarray}
Correcting the $X$ error would require the use of a standard quantum
error correcting code (concatenated with this continuous variable
code). See \cite{Nielsen00} for an introduction to quantum error-correcting
codes.

\begin{figure}
\includegraphics[width=8cm]{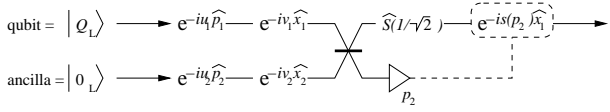}
\caption{Procedure to correct $p$ shifts acting on the qubit.  The circuit shows errors acting on the qubit and ancilla.  Then the qubit and ancilla modes meet in a beam splitter, mode 1 is squeezed, the $p$-quadrature of mode 2 is measured, and a correcting shift of $s(p_2)$ is applied to the qubit. \label{correcting_p}}
\end{figure}

The procedure for correcting shifts to the qubit's $p$-quadrature wave function is similar to that used for $x$ shifts. It is pictured in Fig. \ref{correcting_p}.  In this case we prepare the ancilla qubit in the state $|0_\text{L}\rangle$, we measure the $p$-quadrature basis of mode 2, we apply the squeezing operator $\hat{S}(1/\sqrt{2})$, and the correcting shift is $e^{-is(p_2)\hat{x}_1}$. The errors and correcting procedure produces the new state
\begin{eqnarray}
|Q_\text{L}\rangle & \rightarrow & e^{i\theta(u_1,v_1,u_2,v_2,n)} e^{-i(u_1-u_2)\hat{p}_1} \nonumber \\
& & \times e^{-i\left(v_1-\frac{1}{2}\text{Mod}_{2\sqrt{\pi}}\left(2v_1+2v_2 \right) \right)\hat{x}_1
}|Q_\text{L}\rangle,
\end{eqnarray}
where the phase factor $e^{i\theta(u_1,v_1,u_2,v_2,n)}$ is independent of the initial qubit state. When $|v_1+v_2|<\sqrt{\pi}/2$ the correcting procedure is successful, resulting in the state
\begin{eqnarray}
|Q_\text{L}\rangle & \rightarrow & e^{i\theta(u_1,v_1,u_2,v_2,n)} e^{-i(u_1-u_2)\hat{p}_1} \nonumber \\
& & \times e^{-iv_2\hat{x}_1}|Q_\text{L}\rangle.
\end{eqnarray}
When $v_1$ and $v_2$ are too large, the correcting circuit creates a Pauli $Z$ error in the logical qubit basis.

Armed with a clear understanding of the error correction procedure, we
can formulate a bound on the maximum error that this scheme can always
correct. In the following we imagine that the only error source is in
the preparation of the qubit and ancilla states. We assume that all of
the operations of the error correcting procedure are error free. In
the above analysis we have shown how errors in ancillas are transferred
to the qubit.  We would like to find a bound on the size of shift
errors affecting the qubit and ancillas that ensures that the qubit
never receives an $X$ or $Z$ error and that the sizes of $x$ and $p$
shifts do not grow after multiple steps of error correction.  To
accomplish this we follow the qubit's $x$ and $p$ shifts as it
passes through multiple steps of correction.

Before the error correction procedure, the qubit has errors $u_1$ and
$v_1$, and the first ancilla has errors $u_2$ and $v_2$.  If we first
correct errors of the $x$-quadrature, the qubit will have errors $u_2$
(in $x$) and $v_1-v_2$ (in $p$), provided that
$|u_1+u_2|<\sqrt{\pi}/2$.  This result comes directly from
Eq. (\ref{x_correction}).  Now, we should correct errors of the
$p$-quadrature using a new ancilla, which may have errors $u_3$ and
$v_3$.  The $p$ error correction produces a new qubit with errors
$u_2-u_3$ (in $x$) and $v_3$ (in $p$), provided that
$|v_1-v_2+v_3|<\sqrt{\pi}/2$.  At this point, the first stage of error
correction is complete.  The qubit's original errors have been
entirely eliminated and replaced by errors introduced by the ancillas.
A second correction of the $x$-quadrature using a third ancilla with
errors $u_4$ and $v_4$ is successful if
$|u_2-u_3+u_4|<\sqrt{\pi}/2$.  The qubit now has errors $u_4$ and
$v_3-v_4$.  In this and each subsequent stage, we find that at
any time, the qubit has errors inherited from one ancilla in the
quadrature that was just corrected and errors from two ancillas in
the quadrature, which should be corrected next.  During the next
correction errors from a new ancilla are introduced.  The correction
succeeds if the magnitudes of the now three errors is less than
$\sqrt{\pi}/2$.  Therefore, we can be certain that repeated error
correction steps will be successful if the magnitude of all errors
shifts are smaller than $\sqrt{\pi}/6$.

\begin{figure}
\includegraphics[width=8cm]{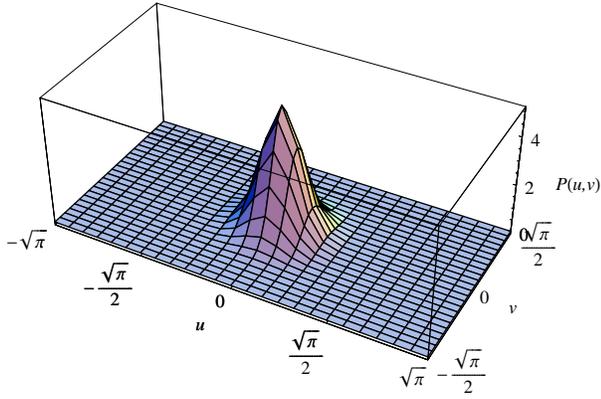}
\caption{$P(u,v)$ the probability density that the $|\tilde{0}\rangle$ approximate qubit state contains errors $u$ (in its $x$-quadrature) and $v$ (in its $p$-quadrature). Here we have used $\Delta=k=1/4$.
\label{Puvzero}}
\end{figure}

\begin{figure}
\includegraphics[width=8cm]{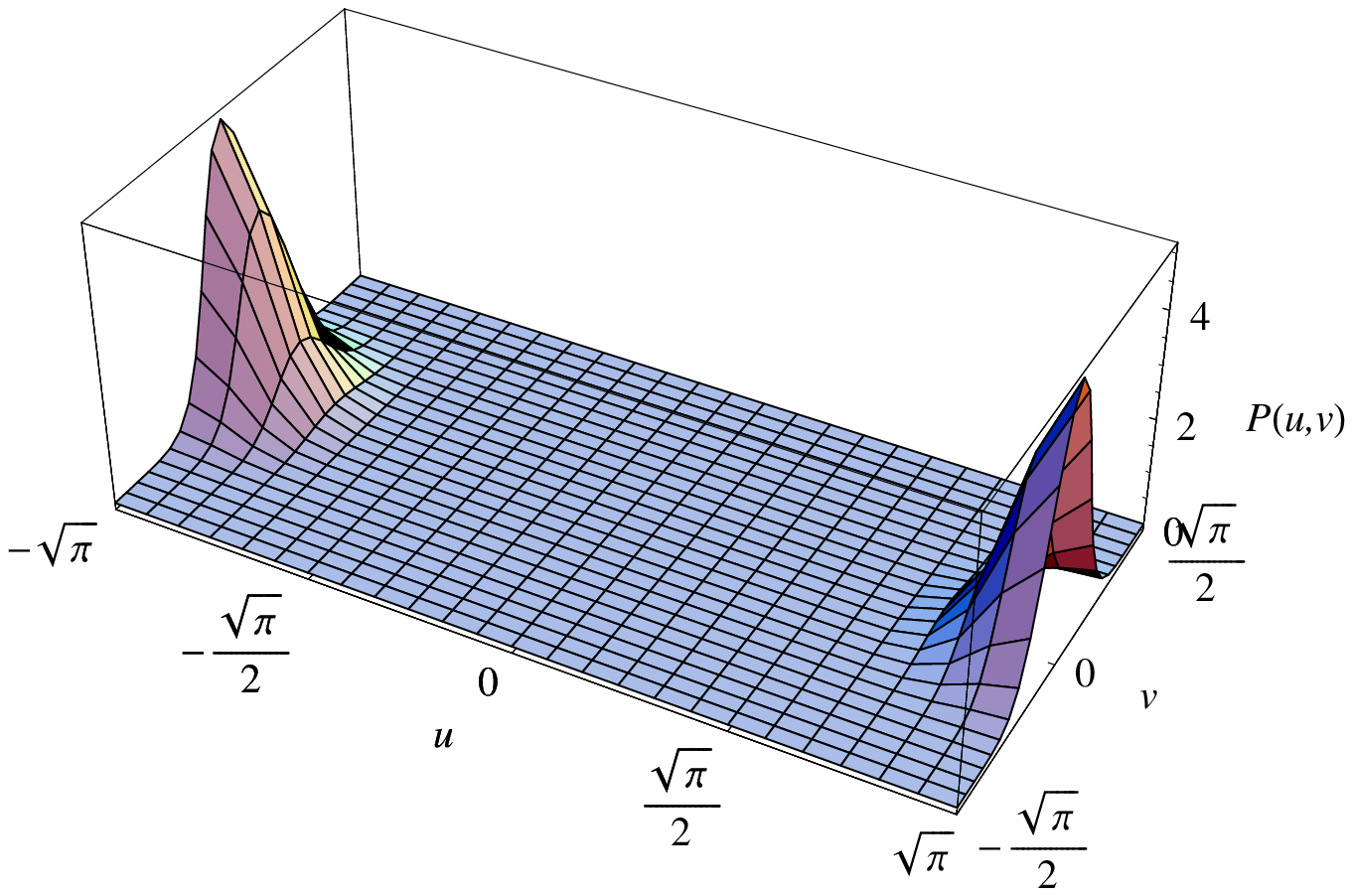}
\caption{$P(u,v)$ the probability density that the $|\tilde{1}\rangle$ approximate qubit state contains errors $u$ (in its $x$-quadrature) and $v$ (in its $p$-quadrature). Here we have used $\Delta=k=1/4$.
\label{Puvone}}
\end{figure}

We would now like to calculate the probability that an approximate qubit state such as $|\tilde{0}\rangle$ or $|\tilde{1}\rangle$ has shifts smaller than $\sqrt{\pi}/6$ in both $x$- and $p$-quadratures.  To find this probability we decompose the approximate qubit states in a basis defined by the states that are $x$- and $p$-shifts of $|0_\text{L}\rangle$, which we may express as
\begin{eqnarray}
|u,v\rangle & = & \pi^{-1/4} e^{-iu\hat{p}}e^{-iv\hat{x}}|0_\text{L}\rangle \\
 & = & \pi^{-1/4} \sum_{s=-\infty}^\infty e^{-iv2s\sqrt{\pi}}|x=2s\sqrt{\pi}+u\rangle.
\end{eqnarray}
We can express any state $|\psi\rangle$ of a single oscillator in this basis using a wave function $f(u,v)=\langle u,v|\psi\rangle$:
\begin{equation}
|\psi\rangle=\int_{-\sqrt{\pi}}^{\sqrt{\pi}}\text{d}u\int_{-\frac{\sqrt{\pi}}{2}}^{\frac{\sqrt{\pi}}{2}}\text{d}vf(u,v)|u,v\rangle.
\end{equation}
The limits of the above integral are chosen to match the periodicity in $|0_\text{L}\rangle$.  The probability density for a state having shifts $u$ and $v$ from $|0_\text{L}\rangle$ is simply
\begin{equation}
P(u,v)=|f(u,v)|^2.
\end{equation}
Examples of $P(u,v)$ are shown in figures \ref{Puvzero} and \ref{Puvone}.

\begin{figure}
\includegraphics[width=8cm]{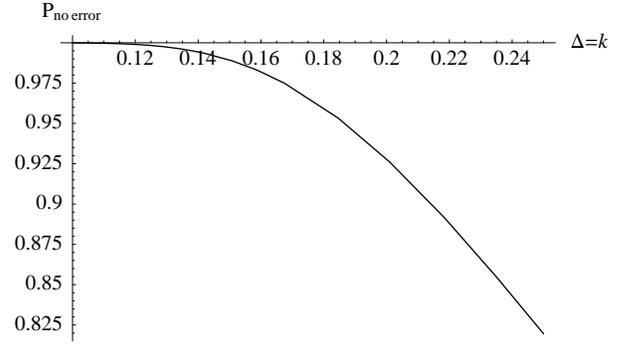}
\caption{Here we plot $P_{\text{no error}}$, the probability that an approximate qubit state has shifts in both $x$- and $p$-quadratures less than $\sqrt{\pi}/6$ as a function of $\Delta=k$.  This shows the case of $|\tilde{0}\rangle$, but the $|\tilde{1}\rangle$ plot is indistinguishable.
\label{PnoerrorGKP}}
\end{figure}

The probability that the $|\tilde{0}\rangle$ state has errors less than $\sqrt{\pi}/6$ (which guarantees that its errors may always be corrected) is \begin{eqnarray}
P_{\text{no error}}= \int_{-\frac{\sqrt{\pi}}{6}}^{\frac{\sqrt{\pi}}{6}}\text{d}u
\int_{-\frac{\sqrt{\pi}}{6}}^{\frac{\sqrt{\pi}}{6}}\text{d}v
P(u,v).
\end{eqnarray}
We plot $P_{\text{no error}}$ in Fig. \ref{PnoerrorGKP}.  Achieving $P_{\text{no error}}=0.9$ requires $\Delta=k=0.214$, and $P_{\text{no error}}=0.99$ requires $\Delta=k=0.149$.

\begin{figure}
\includegraphics[width=8cm]{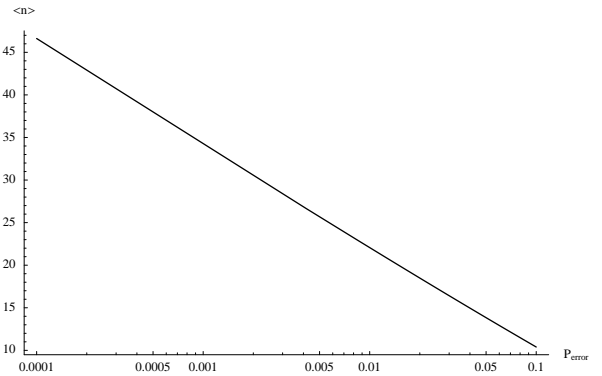}
\caption{Here we plot the mean number of photons contained in an approximate qubit state as a function of the probability that such a state has a shift error larger than $\sqrt{\pi}/6$.
\label{nvsPerror}}
\end{figure}

We generally believe that protecting an oscillator from decoherence usually becomes more difficult when the oscillator contains large numbers of photons, so we are motivated to calculate the mean number of photons in the approximate qubit states.  A crude estimate of this quantity for small $\Delta$ and $k$ is given by
\begin{eqnarray}
\langle n\rangle \sim \frac{1}{4\Delta^2}+\frac{1}{4 k^2}.
\end{eqnarray}
In Fig. \ref{nvsPerror} we plot the mean number of photons contained in an approximate qubit state as a function of $P_{\text{error}}=1-P_{\text{no error}}$ using an exact expression.  We find that an approximate qubit state with $P_{\text{no error}}=0.9$ must contain $\langle n \rangle = 10.4$ photons, and to achieve $P_{\text{no error}}=0.99$ requires $\langle n \rangle = 22.1$ photons.

Production of these states is likely to be very challenging. We are
aware of four proposals for production of qubit states for the GKP
quantum computer. The first is from the original work of GKP
\cite{Gottesman01}, in which they propose the use of a two-mode
Hamiltonian of the form $\hat{x}_1\hat{n}_2$. This might be achieved
by coupling an optical mode with a mirror that may exist in a
coherent superposition of position states. For a discussion of the
possibility of such experiments see \cite{Giovannetti00, Pinard05} and
their references. The second proposal is from Travaglione and Milburn
\cite{Travaglione02}. They describe a method that prepares the qubit
states in the oscillatory motion of a trapped ion rather than the
photons in an optical mode. The third proposal
is by Pirandola {\em et al.}  \cite{Pirandola04} and discusses the
preparation of optical GKP states using a two mode Kerr interaction (described by a Hamiltonian of the form $\hat{n}_1\hat{n}_2$) followed by a homodyne measurement of one of the modes.  The same authors also describe a fourth method for GKP state production in \cite{Pirandola05a, Pirandola05b}.
This method would prepare the GKP state in the motion of a neutral atom that
interacts with a photon field confined to a high finesse optical
cavity.  They show that this may be done so that
the atom is trapped in the cavity, or the atom may be free to pass
through the cavity.  State preparation is currently the most
problematic aspect of the GKP quantum computation scheme, and it is an
area in need of more experimental effort.

In this paper we have investigated error correction in the GKP quantum
computer, considering errors in both qubit and ancilla states. Provided
that all input states have displacement errors in their $x$- and
$p$-quadratures less than $\sqrt{\pi}/6$, the error correction schemes
will operate successfully. Because it is impossible to prepare
perfectly error free GKP states, we have calculated the probability
that approximate GKP states contain an error larger than
$\sqrt{\pi}/6$. An approximate GKP state with error probability less
than $0.1$ must contain a mean number of photons greater than $10.4$.

There are some cases in which states with errors larger than
$\sqrt{\pi}/6$ will not cause logical qubit errors. For example, the
qubit's and the ancilla's errors may actually cancel one
another. Because of effects like this, our $\sqrt{\pi}/6$ bound may be
refined with more detailed analysis. We have not considered full fault
tolerant computation in the presence of noisy logic operations, nor
have we considered the effects of phase errors or photon
absorption. Phase errors and photon absorption are likely to be
primary error sources for this (or any) optical quantum computer.
Because linear superpositions of displacements span the space of
possible single qubit errors, one may correct phase and absorption
errors using the GKP error correction scheme. However, phase errors
affecting states with large numbers of photons correspond to large
displacements.  This fact highlights two competing influences: (1) To
lower the number of intrinsic errors in each approximate qubit state,
we must prepare states with large numbers of photons. (2) To reduce
the displacements caused by phase errors, we should prepare states with
fewer photons.  A detailed analysis of these constraints requires
further study.

{\em Acknowledgments}
We thank Hilma Vasconcelos and Rich Mirin for helpful comments on the manuscript.

\end{document}